\documentclass{article}

\usepackage[right= 0.75in, left= 0.75in, top=1.05in, bottom=1.15in]{geometry}

\usepackage{lineno,hyperref,natbib}
\modulolinenumbers[5]

\usepackage[utf8]{inputenc} 

\usepackage{amsmath}
\usepackage{amsfonts}
\usepackage{mathrsfs} 
\usepackage{booktabs} 
\usepackage{array} 
\usepackage{color}
\usepackage{graphicx}
\usepackage{paralist} 
\usepackage{verbatim} 
\usepackage{subfig} 
\usepackage{bm}
\usepackage[affil-it,auth-sc]{authblk}

\DeclareMathOperator{\divg}{\mbox{div}} 
\DeclareMathOperator{\divgs}{\mbox{div}_s} 
\DeclareMathOperator{\grads}{\nabla_{\!s}} 

\newcommand{\mb}[1]{\mathbf{#1}}
\newcommand{\rr}[1]{\textcolor{black}{#1}}
\newcommand{\mc}[1]{\mathcal{#1}} 
\newcommand{\mbb}[1]{\mathbb{#1}} 
\newcommand{\wt}[1]{\widetilde{#1}} 
\newcommand{\wh}[1]{\widehat{#1}} 

\author[1, \footnote{Corresponding author -- e-mail address:\,\texttt{alucanto@sissa.it}.}]{Alessandro Lucantonio}
\author[2]{Luciano Teresi}
\author[1]{Antonio DeSimone}
\affil[1]{\small{SISSA - International School for Advanced Studies, via Bonomea 265, 34136 Trieste - Italy}}
\affil[2]{\small{Department of Mathematics and Physics, Università Roma Tre, via della Vasca Navale 84, 00146 Roma - Italy}}

\title{Continuum theory of swelling material surfaces with applications to thermo-responsive gel membranes and surface mass transport}

\date{}

\begin{document}

\maketitle

\begin{abstract}
Soft membranes are commonly employed in shape-morphing applications, where the material is programmed to achieve a target shape upon activation by an external trigger, and as coating layers that alter the surface characteristics of bulk materials, such as the properties of spreading and absorption of liquids. In particular, polymer gel membranes experience swelling or shrinking when their solvent content change, and the non-homogeneous swelling field may be exploited to control their shape. Here, we develop a theory of swelling material surfaces to model polymer gel membranes and demonstrate its features by numerically studying applications in the contexts of biomedicine, micro-motility, and coating technology. We also specialize the theory to thermo-responsive gels, which are made of polymers that change their affinity with a solvent when temperature varies.
\end{abstract}

\section{Introduction}

Among soft active materials, \textit{i.e.}~materials that respond with a mechanical deformation to a non-mechanical stimulus (electrical field, exposure to a solvent, pH change, temperature field), polymer gels play a major role in the current research on novel micro- and nano-devices. The mechanical characteristics of these materials closely resemble those of biological tissues and thus make them candidates for biomedical applications and bio-inspired devices \citep{Ottenbrite}. 

In many applications, gels are employed in the form of membranes \citep{Stuart2010, Ionov2011}; in particular, in self-shaping materials, these membranes can undergo prescribed three-dimensional shape transformations, by exploiting suitable spatial modulations of the local degree of swelling. The non-uniform swelling field may be obtained through a non-homogeneous in-plane \citep{Klein2007, Kim2012a, Wu2013} or through-the-thickness \citep{Hu1998, Sawa2010, Lucantonio08112014} distribution of the cross-linking density, or by a localized exposure of the system to a solvent \citep{Holmes2011, Lucantonio2012, Pandey2013}. Alternatively, the shape of a swelling membrane may be manipulated by harnessing the multiphysics coupling between elasticity and solvent transport, specifically through the combination of solvent stimulation with an applied pre-stretch \citep{Lucantonio2014, Lucantonio2014a}. Apart from shape-morphing applications, polymer gels have been most successfully employed as drug delivery systems over the past few decades \citep{Hoare2008}. Among the many designs that have been proposed, in reservoir systems the drug core is confined by a spherical gel membrane, which is occasionally made of a stimuli-responsive material, such as a thermo-responsive gel, in order to achieve a pulsatile drug delivery \citep{Peppas2000, Kikuchi2002}. 

Motivated by these applications in diverse and emerging fields, here we study a polymer gel membrane that undergoes swelling when exposed to a solvent. We model such a membrane as a \textit{swelling material surface}, an extension to swelling materials of the concept of material surface, which dates back to \cite{Gurtin1975} and involves, in general, a surface endowed with a physical structure ruled by a set of balance equations (balance of mass, forces, moments, energy, \dots). In particular, we model the coupled solvent transport and elasticity of the polymer network, and introduce a thickness microstructural variable that accounts for the volume change of the membrane caused by the absorption of solvent. In \cite{McBride2011} a nonlinear continuum thermomechanics formulation that accounts for surface structures and includes the effects of diffusion and viscoelasticity was presented, and afterwards numerically implemented \citep{Javili2014}. Other relevant works where the theory of material surfaces has been extended to include surface mass transport are \cite{Ganghoffer2005, Steinmann2012}. However, swelling has not been considered, and a very limited number of applications has been presented, none concerning soft active materials, in general, or polymer gels, in particular.
Here, we focus on coating gel membranes, that is, we model swelling material surfaces that cover boundaries of bodies (boundary material surfaces), and study several systems with relevant applications in the contexts of biomedicine and micro-motility. Specifically, boundaries of homogeneous gels deserve particular attention, as they exhibit transport properties that differ from those of the bulk material, thus affecting surface phenomena, such as spreading and absorption of liquids \citep{Starov2002}. The theory is sufficiently general to be applicable to stand-alone soft membranes, even in the absence of swelling.

We adopt a direct approach in the formulation of the governing equations for the material surface, instead of deducing them from a three-dimensional theory. Precisely, we prescribe a virtual work functional and use it as a tool to derive the local balance of forces and moments, while the balance of solvent mass for the surface is directly stated in integral form and then localized. A deductive approach, instead, is employed for the swelling constraint, which relates the solvent volume fraction to the volume change of the membrane, and for the derivation of the surface free energy, because both involve the notion of volume change, which pertains to a three-dimensional body. Eventually, we obtain a thermodynamically consistent theory that fits in the theoretical framework for swelling gels set in \cite{AL2013}. 

The paper is organized as follows. In Section~\ref{sec:preliminaries}, we set the notation, the kinematics and recall several definitions from differential geometry and calculus on surfaces. In Section~\ref{sec:balanceeq}, we collect the governing equations for a three-dimensional body with a boundary material surface, both subject to swelling, together with the kinematic constraints that relate swelling to solvent uptake. In Section~\ref{sec:consteq}, we deal with thermodynamical issues and specify the representation forms for the free energy of the body and the swelling material surface that are suitable for the study of gels. In Section~\ref{sec:weakform}, we present the weak formulation of the governing equations of the model, in order to enable their implementation in a finite element software. Finally, in Section~\ref{sec:applications}, we discuss applications of the theory to a smart drug delivery system, a temperature-activated gel micro-crawler, and to a coated gel where a competition between surface spreading and absorption of a liquid occurs.

\section{Preliminaries: notation and kinematics} 
\label{sec:preliminaries}
We consider, as a reference scenario, a soft membrane that is swollen with a liquid solvent and lies on the boundary of a three-dimensional body, also made of a soft, swellable material.  Both the membrane and the body undergo swelling or shrinking when their solvent content changes. 
We model the (three-dimensional) membrane as a material surface with a scalar microstructure that measures its thickness variation.  
We assume that the \rr{reference configuration} $\mc{S}\subseteq \partial\mc{B}$ of the material surface partially (or totally) covers the boundary $\partial \mc{B}$ of the \rr{reference configuration} $\mc{B} \subset \mc{E}$ of the body, where $\mc{E}$ is the three-dimensional Euclidean space whose translation space is $\mc{V}$. Elements (material points) of the sets $\mc{B}$ and $\mc{S}$ will be labelled as $X$. 

Upon introducing the time $t \in \mc{I} \subset \mbb{R}$, we denote by $f: \mc{B} \times \mc{I} \rightarrow \mc{E}$ the motion of the body, which is a one-parameter family of (smooth injective) deformation mappings such that $x = f(X,t)\in\mc{E}$. We assume that the boundary material surface is always bonded to the body, so that the motion $f_s:\mc{S}\times\mc{I}\rightarrow\mc{E}$ of the surface is given by the restriction $f|_{\mc{S}}$ of the motion $f$. The images $\mc{B}_t = f(\mc{B},t)$ and $\mc{S}_t = f_s(\mc{S},t)$ of $\mc{B}$ and $\mc{S}$ are the current configurations of $\mc{B}$ and $\mc{S}$ at time $t$, respectively. A superposed dot denotes differentiation with respect to time, which is regarded as a parameter for the equation of balance of forces, under the hypothesis of negligible inertial forces. Related to the motions of the body and the surface, we define the displacement fields $\mb{u}(X,t) = f(X,t)-X, X\in\mc{B}$ and $\mb{u}_s(X,t) = f_s(X,t) - X = \mb{u}|_{\mc{S}}, X\in\mc{S}$.

\begin{figure}[!h]
\centering
\includegraphics[scale=0.85]{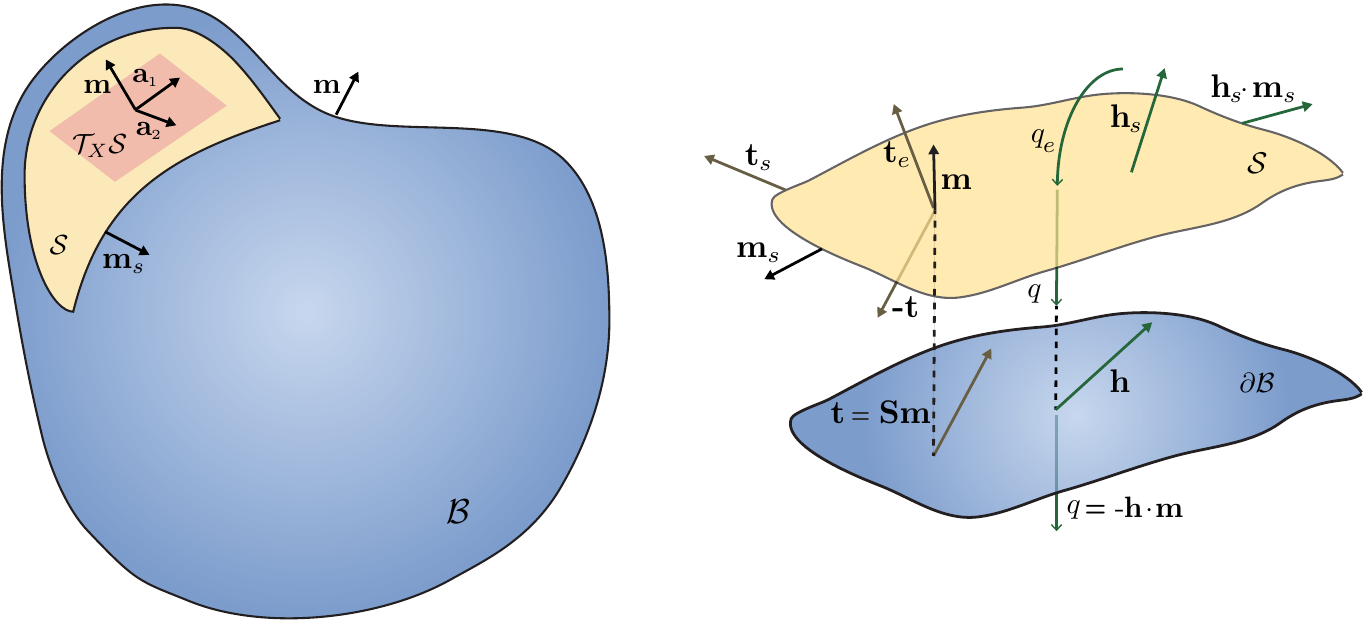}
\caption{Sketch of the reference configuration of a body with \rr{a swelling material surface on part of its boundary}. The exploded view shows the notation and the directions for the fluxes and the contact forces exchanged between the body and the material surface.}
\label{fig:sketchconf}
\end{figure}

\rr{Let us indicate with $\mc{T}\mc{S}$ the tangent bundle of the surface $\mc{S}$; its current counterpart is denoted by $\mc{T}\mc{S}_t$.} 
\rr{The tangent space $\mc{T}_X\mc{S}$ is spanned by the covariant basis $\mb{a}_\alpha(X)$ and the contravariant basis $\mb{a}^\alpha(X)$, with $\alpha=1,2$ (Figure~\ref{fig:sketchconf}), while $\mc{T}_x\mc{S}_t$ is spanned by the covariant basis $\mb{g}_\alpha(x,t)$ and the contravariant basis $\mb{g}^\alpha(x,t)$. We then introduce the surface projection field \rr{$\mb{P}$ such that} $\mb{P}^T\mb{P}  = \mb{I} - \mb{m}\otimes\mb{m}$, where $\mb{m} = \mb{a}_1 \times \mb{a}_2 / |\mb{a}_1 \times \mb{a}_2|$ is the unit normal to an area element of the reference surface and $\mb{I}$ is the identity of $\mc{V}$. The tensor $\mb{P}(X)$ projects a vector that belongs to $\mc{V}$ on $\mc{T}_X\mc{S}$, whereas its transpose, which is called inclusion, performs the inverse transformation.} 
\rr{Analogously, we can introduce the current surface projection $\mb{P}_t$ such that $\mb{P}_t^T\mb{P}_t = \mb{I}-\mb{n}\otimes\mb{n}$, where $\mb{P}_t^T$ is the associated inclusion, and $\mb{n}=\mb{g}_1 \times \mb{g}_2 / |\mb{g}_1 \times \mb{g}_2|$ is the outward normal to $\mc{S}_t$.} 

\rr{We denote by $\grads$ and $\divgs$ the surface gradient and the surface divergence, respectively, of a field defined over $\mc{S}$ \cite{Gurtin1975, Murdoch1990}}.
By means of these operators, we can represent the tangent maps (also called the deformation gradients) $\mb{F} = \nabla f = \mb{I} + \nabla\mb{u}$ and $\mb{F}_s  = \mb{P}_t\mb{F}|_{\mc{S}}\mb{P}^T$  that send, locally and at any time, elements of $\mc{T}_X\mc{B}$ and $\mc{T}_X\mc{S}$  to $\mc{T}_x\mc{B}_t$ and $\mc{T}_x\mc{S}_t$, respectively. 
In the following, we will make use of the non-invertible tensor $\wh{\mb{F}} = \grads f_s =  \mb{P}^T + \grads{\mb{u}}_s = \mb{F}|_\mc{S}\mb{P}^T$, whose invertible counterpart is $\mb{F}_s$. As measures of stretch, we take the right Cauchy-Green tensors $\mb{C} = \mb{F}^T\mb{F}$ and $\mb{C}_s = \wh{\mb{F}}^T\wh{\mb{F}} = \mb{F}_s^T\mb{F}_s$.

As usual, the determinant $J=\det\mb{F}$ of the deformation gradient measures the local change in volume, while the local area change is computed through the cofactor $\mb{F}^\star = J\mb{F}^{-T}$ as $|\mb{F}^\star\mb{m}|$.
\rr{Analogously, the ratio $\mbox{d}S_t/\mbox{d}S$ between the current and reference area elements of the material surface is measured by  $|\mb{F}^\star|_{\mc{S}}\mb{m}| = J_s = \det \mb{F}_s$}.

Because solvent absorption locally induces a change in volume of the boundary membrane modeled by the material surface, we attach to $\mc{S}$ a \textit{scalar microstructure} $\delta(X,t)$, $X \in \mc{S}$ that measures the thickness stretch of the membrane, \textit{i.e.}~the ratio between the current and the reference thickness $h(X)$ of the membrane. \rr{The kinematics of the \textit{swelling material surface} $\mc{S}$ is thus described by the pair of fields $(\mb{u}_s,\delta)$.} The thickness stretch $\delta$ contributes to the volume change of the membrane together with the area change of the surface measured by $J_s$. We will specify the relation between solvent uptake and volume change of the membrane in Section~\ref{sec:swellingconstraints}. 

\section{Balance equations for a body with a swelling material surface}
\label{sec:balanceeq}
We consider elasticity and solvent transport as the physics that characterize the behavior of both the body and the boundary membrane. 
To account for the swelling of the body and of the boundary membrane, we formulate the kinematic constraints that express the volume change induced by the migration of solvent. \rr{We then arrive at the formulation of a swelling material surface model.} Finally, we establish the constitutive equations for the body and the swelling material surface through thermodynamical arguments.

\subsection{Balance equations for the body}

With reference to \cite{AL2013}, we briefly recall the mechanical balance equations and the balance of solvent mass for the body. We consider a system of forces $(\mb{f},\mb{t})$ acting on the body, with $\mb{f}$ the body load per unit reference volume and $\mb{t}$ the boundary load per unit reference area. As usual, the balances of forces and moments for the body read
\begin{align}
\divg{\mb{S}} + \mb{f} = \mb{0} \quad \mbox{on}\ \mc{B}\times\mc{I}\,, && \mb{t} = \mb{S}\mb{m} \quad \mbox{on}\ \partial\mc{B}\times\mc{I}\,, && \mbox{skw}(\mb{SF}^T) = \mb{0} \quad \mbox{on}\ \mc{B}\times\mc{I}\,, \label{eq:bala3d}
\end{align}
where $\mb{S}$ is the reference (Piola-Kirchhoff) stress. In particular, on $\mc{S}\subset\partial \mc{B}$, $\mb{t}$ is the contact force per unit area applied by the membrane to the body. 
 
For solvent transport, upon introducing the solvent concentration field $c: \mc{B}\times\mc{I}\rightarrow \mbb{R}^{+}$ per unit reference volume and the solvent mass flux $\mb{h}:\mc{B}\times\mc{I}\rightarrow\mc{V}$ per unit reference area, the balance of solvent mass reads
\begin{align}
\dot{c} = -\divg{\mb{h}} \quad \mbox{on}\ \mc{B}\times\mc{I}\,, && -\mb{h}\cdot\mb{m} = q \quad \mbox{on}\ \partial\mc{B}\times\mc{I}\,,
\end{align}
where $q$ is the solvent mass boundary source, which corresponds on $\mc{S}$ to the flux from the membrane to the body.
\subsection{Balance equations for the swelling material surface}

In a similar fashion to what we have done for the body, we describe the forces acting on an arbitrary, regular subsurface $\mc{P}_s \subset \mc{S}$ by the pair of vector fields $(\mb{f}_s,\mb{t}_s)$, where $\mb{f}_s$ is the force per unit area, which represents the distributed surface load over $\mc{P}_s$, and $\mb{t}_s$ is the contact force per unit length along $\partial \mc{P}_s$. The surface load $\mb{f}_s$ consists of the distributed contact forces $-\mb{t}$ exerted by the body on the  material surface and of the distributed applied load $\mb{f}_e$ exerted by the environment on $\mc{S}$: $\mb{f}_s = \mb{f}_e -\mb{t}$. Additionally, we introduce the surface load $c_n$ that represents the external force per unit area spending power on changes in thickness. The outward unit normal to $\partial \mc{P}_s$ will be indicated with $\mb{m}_s$.

We proceed with deriving the balance equations following the method of virtual power \citep{Germain1973}. Thus, we prescribe the following version of the principle of virtual power
\begin{align}
\label{eq:virtpower}
\int_{\mc{P}_s}{(\mb{S}_s\cdot\grads\wt{\mb{v}}_s + \sigma_n \wt{\eta})} = \int_{\partial\mc{P}_s}{\mb{t}_s\cdot\wt{\mb{v}}_s} + \int_{\mc{P}_s}{(\mb{f}_s\cdot\wt{\mb{v}}_s + c_n\wt{\eta})}
\end{align}
for any part $\mc{P}_s$, and for any choice of the virtual surface velocity field $\wt{\mb{v}}_s$ and of the virtual thickness stretch rate $\wt{\eta}$.
Here we have introduced the stresses $\mb{S}_s:\mc{T}\mc{S}\times\mc{I}\rightarrow\mc{V}$ and $\sigma_n:\mc{S}\times\mc{I}\rightarrow\mbb{R}$ and we have neglected both applied and contact couples, consistent with the classical theory of membranes (see Remark below). The functional at the right hand side is the external virtual power $\Pi_{\rm ext}$, while that at the left hand side is the internal virtual power $\Pi_{\rm int}$.

First, by taking $\wt{\eta}=0$ in eq.~\eqref{eq:virtpower}, using the surface divergence theorem for a tensor field \citep{Gurtin1975} and localizing, we obtain the equations of balance of forces for the surface
\begin{align}
\label{eq:balanceforces}
&\divgs{\mb{S}_s} +\mb{f}_s = \mb{0}\,, && \mbox{on}\ \mc{S}\times\mc{I}\,, \\
&\mb{t}_s = \mb{S}_s\mb{m}_s\,, && \mbox{on}\ \partial\mc{S}\times\mc{I}\,.
\end{align}
Then, with $\wt{\mb{v}}_s = \mb{0}$, from eq.~\eqref{eq:virtpower} we derive the balance of thickness forces on $\mc{S}\times\mc{I}$
\begin{align}
\label{eq:balathick}
\sigma_n = c_n\,.
\end{align} 
To give a physical interpretation to eq.~\eqref{eq:balathick}, we may picture the thickness microstructure as a distribution of deformable segments attached to the surface, which only resist to stretching; therefore, eq.~\eqref{eq:balathick} represents the force balance along \rr{the} direction parallel to each segment and $\sigma_n$ plays the role of a thickness stress.

According to the principle of frame-indifference, the internal virtual power must be invariant under changes of observer. 
\rr{This requirement provides}, on localizing, 
\begin{align}
\label{eq:balancemoments}
\mbox{skw}(\wh{\mb{F}}\mb{S}_s^T) = \mb{0}\,,
\end{align}
or, equivalently, $\mb{F}_s\mb{S}_s^T\mb{P}_t^T = \mb{P}_t\mb{S}_s\mb{F}_s^T$. Equation~\eqref{eq:balancemoments} is a restriction on the constitutive prescriptions for $\mb{S}_s$ and implies that the surface does not carry internal forces along the normal direction to its deformed shape; indeed\footnote{Unless explicitly specified, summation convention is used, with Greek indices ranging from 1 to 2.}, \rr{$\mb{S}_s\mb{a}^\alpha \cdot \mb{n} =  (\mb{S}_s\mb{a}^\beta \cdot \mb{n})\mb{g}_\beta \cdot \mb{g}^\alpha = \mbox{skw}(\wh{\mb{F}}\mb{S}_s^T)\mb{n}\cdot\mb{g}^\alpha = 0$.} 

As concerns surface solvent transport, we introduce the \rr{surface} solvent concentration $c_s: \mc{S} \times \mc{I} \rightarrow \mbb{R}^{+}$, which measures the number of solvent moles per unit area of $\mc{S}$. The material surface can exchange solvent with the bulk $\mc{B}$ pointwise in $\mc{S}$, and with the environment through its boundary $\partial \mc{S}$. Hence, the balance of solvent mass for $\mc{P}_s \subset \mc{S}$ reads
\begin{align}
\label{eq:massbalsurf}
\frac{\mbox{d}}{\mbox{d}t} \int_{\mc{P}_s}{c_s} = -\int_{\partial \mc{P}_s}{\mb{h}_s \cdot \mb{m}_s} + \int_{\mc{P}_s}{q_s}\,,
\end{align}
where $\mb{h}_s: \mc{S} \times \mc{I} \rightarrow \mc{T}\mc{S}$ is the \rr{surface} solvent flux and the solvent mass source $q_s = -q + q_e$ includes the exchange of solvent with the bulk and the external supply $q_e$ of solvent.
Using the \rr{surface divergence theorem \cite{Gurtin1975} for a vector field}, eq.~\eqref{eq:massbalsurf} localizes to
\begin{align}
\label{eq:localmassbalsurf}
\dot{c}_s = -\divgs{\mb{h}_s} + q_s\,, && \mbox{on}\ \mc{S}\times\mc{I}\,.
\end{align}
\paragraph{Remark} Equations \eqref{eq:balanceforces} and \eqref{eq:balancemoments} are the governing equations for a membrane, \textit{i.e.}~a thin shell-like body with negligible bending stiffness. These equations may also be recovered from the model of special Cosserat shell \citep{AntmanBook}, where the director $l$ is constrained to be a unit vector (the thickness does not change), by imposing that the contact couples, represented\footnote{Here we use the same notation as in \citep{DiCarlo2001}.} by the couple tensor $M$, the applied body couples $c\times l$ and the boundary couples $m\times l$ vanish. For shells with thickness distension \citep{DiCarlo2001},  the following balance of director forces is added to the governing equations:
\begin{align}
&(\divgs{M}-Ne+c)\cdot l = 0\,, && \mbox{on}\ \mc{S}\,, \label{eq:balADC}\\
&M\nu\cdot l = m\cdot l\,, && \mbox{on}\ \partial\mc{S}\,,
\end{align}
where $\nu$ is the unit normal to $\partial\mc{S}$.
Then, eq.~\eqref{eq:balADC} reduces to eq.~\eqref{eq:balathick} under the hypotheses $M = 0$ and $m\cdot l=0$, and upon recognizing that: \textit{i}) $\sigma_n$ corresponds to the component $Ne\cdot l$ along the director $l$ of the contact force per unit length $Ne$ exchanged between parts of the shell through cut-planes that are orthogonal to the reference director $e$; \textit{ii}) $c_n$ corresponds to the director bulk-force $c\cdot l$. In turn, the balance equations for a shell with thickness distension may be derived from the principle of virtual power for a three-dimensional Cauchy continuum, by employing the appropriate representations for the virtual displacement fields. Such a deductive approach allows to extend the present model to a richer kinematics, and to derive a hierarchy of structural theories having different degrees of approximation \rr{with respect to a parent three-dimensional model}. Here, however, we favor a direct approach, since it avoids the introduction of additional structure (\textit{i.e.}~the director) that is redundant \rr{for the swelling material surface model}.

\subsection{Swelling constraints}
\label{sec:swellingconstraints}
Up to this point, we have introduced the balance equations that allow for the description of coupled elasticity and solvent transport phenomena in boundary material surfaces. Henceforth, we specialize the theory to swelling materials, with specific reference to polymer gels, \rr{mixtures of an elastomeric matrix and a fluid where the change in solvent content causes a change in volume of the aggregate}. Because gels often possess a certain amount of solvent in their preparation state, it is convenient to measure volume changes starting from a swollen configuration, where the solvent concentration is homogeneous and equal to $c_{\rm o}$ in the bulk, and to $c_{s\rm o}$ on the surface. The swollen reference configurations $\mc{B}$ and $\mc{S}$ are conceived as being reached from the corresponding dry states through the homogeneous and isotropic deformations characterized by the stretches $\lambda_{\rm o}$ and $\lambda_{s\rm o}$, respectively. Moreover, we specify that $\mc{B}$ and $\mc{S}$ are stress-free and that both the body and the surface are in chemical equilibrium with an external solvent. We will see in Section~\ref{sec:consteq} how to characterize the reference configurations from the viewpoint of thermodynamical equilibrium.

Usually, the solvent is liquid and together with the elastomeric matrix they are assumed to be incompressible, so that the local volume ratio $J$ from $\mc{B}$ relates to the change in solvent concentration as \citep{AL2013}
\begin{align}
\label{eq:swellingconstr}
J = 1+\Omega (c-c_{\rm o})\,,
\end{align}
where $\Omega$ is the solvent molar volume.

For a boundary membrane that consists of a swelling material, we may formulate an analogous constraint. \rr{To this aim, as noted in the introduction,} the swelling constraint involves both the area change of $\mc{S}$ and the thickness change $\delta$ of the membrane. Upon approximating the solvent concentration $c$ within the membrane with its value $c|_{\mc{S}}$ on $\mc{S}$, the solvent volume contained in an infinitesimal reference volume of the membrane $\mbox{d}V = h\,\mbox{d}S$ is $\Omega c|_{\mc{S}} \mbox{d}V = \Omega c_s \mbox{d}S$. \rr{Hence, from eq.~\eqref{eq:swellingconstr}, 
we can express the local volume ratio $J$ for the membrane as}
\begin{align}
\label{eq:swellingconstrsurf}
J = 1 + \Omega \, (c|_{\mc{S}}-c_{\rm o}|_{\mc{S}}) = 1 + \frac{\Omega}{h}\,(c_s-c_{s\rm o})\,,
\end{align}
where $c_{s\rm o} = h\,c_{\rm o}|_{\mc{S}} = (\lambda_{s\rm o}^3-1)h/(\Omega \lambda_{s\rm o}^3)$ is the solvent concentration in the reference state.
By introducing the equivalent molar volume $\Omega_s = \Omega/h$, and by employing the following representation for the deformation gradient of the membrane
\begin{align}
\label{eq:flayer}
\mb{F} \approx \mb{F}|_{\mc{S}} = \wh{\mb{F}}\mb{P}+\delta \mb{n}\otimes\mb{m}
\end{align}
so that $J \approx J|_{\mc{S}} = \delta J_s$, the counterpart of the swelling constraint \eqref{eq:swellingconstr} for the membrane reads
\begin{align}
\label{eq:areaconstraint}
\delta J_s = \delta ( \det \mb{C}_s)^{1/2} =  1 + \Omega_s (c_s-c_{s\rm o})\,.
\end{align}
Notice that, by taking the time derivative of eq.~\eqref{eq:areaconstraint} and replacing $\dot{c}_s$ with eq.~\eqref{eq:localmassbalsurf}, we obtain an evolution equation for the thickness stretch:
\begin{align}
\label{eq:thicknessevo}
\dot{\delta}J_s + \delta\dot{J}_s + \Omega_s (\divgs{\mb{h}_s} - q_s) = 0\,.
\end{align}
The thickness stretch $\delta$ may then be used as a primary variable of the formulation, instead of the boundary concentration $c_s$ -- therefore, eq.~\eqref{eq:thicknessevo}  replaces eq.~\eqref{eq:localmassbalsurf} as the balance of solvent for the surface -- by systematically using the volume constraint \eqref{eq:areaconstraint} to eliminate $c_s$ from the governing equations of the problem.

\subsection{Continuity conditions}

We recall that the displacement field is continuous $\mb{u}_s = \mb{u}|_{\mc{S}}$ because of the continuity of the motion $f$ up to $\partial\mc{B}$, since we are assuming that the boundary material surface is always bonded to the body. In addition, we assume that the \rr{solvent within the} material surface is in chemical equilibrium with that contained in the adjacent layer of material belonging to the body. Hence, at any time, the chemical potential is continuous at $\mc{S}$:
\begin{align}
\label{eq:contcond}
\mu_s = \mu|_{\mc{S}}\,.
\end{align}
This constraint determines (implicitly) the solvent flux $q$ exchanged between the surface and the body and will be enforced through a Lagrange multiplier in the numerical model.
\paragraph{Remark} In \cite{McBride2011} it is shown that the continuity condition \eqref{eq:contcond} need not be assumed from the onset, but it is rather one of the possible ways to satisfy the dissipation inequality (see Section~\ref{sec:consteq}).

\section{Thermodynamics and constitutive equations}
\label{sec:consteq}
With reference to \cite{AL2013}, we assume that the bulk material \rr{(within the body)} is allowed to exchange  with the exterior both mechanical power and \rr{(chemical)} power associated to solvent transport. Thermodynamical arguments based on the Colemann--Noll procedure lead, in an isothermal setting, to the following constitutive restrictions:
\begin{align}
\mb{S} =  \frac{\partial \psi}{\partial \mb{F}}-p\mb{F}^\star\,, && \mu = \frac{\partial \psi}{\partial c}+\Omega p\,, && \mb{h} \cdot \nabla \mu \leq 0\,, \label{eq:constbody}
\end{align}
where $\psi$ is the Helmholtz free energy per unit volume of $\mc{B}$, $\mu$ is the bulk solvent chemical potential and $p$ is the bulk solvent pressure. 
For polymer gels, the Flory-Rehner free energy is commonly employed as a representation form for $\psi$:
\begin{align}
\label{eq:flory3d}
\psi(\mb{F},c) = \frac{1}{2}\frac{G}{J_{\rm o}}(\lambda_{\rm o}^2\,\mb{F}\cdot\mb{F}-3) + \frac{1}{J_{\rm o}} \frac{\mc{R}T}{\Omega}\left[\Omega J_{\rm o} c\log\left(\frac{\Omega J_{\rm o} c}{1+\Omega J_{\rm o} c}\right)+\chi \frac{\Omega J_{\rm o} c}{1+\Omega J_{\rm o} c} \right]\,,
\end{align}
where $G$ is the shear modulus of the dry polymer, $J_{\rm o} = \lambda_{\rm o}^3$ is the initial swelling ratio, $\mc{R}$ is the universal gas constant, $T$ is the absolute temperature and $\chi$ is the dimensionless measure of the solvent-polymer enthalpy of mixing. The first term in eq.~\eqref{eq:flory3d} represents the change in free energy due to the deformation of the polymer network, while the second term is the contribution to the free energy due to the mixing between solvent and polymer network.
Further, \rr{to satisfy eq.~\eqref{eq:constbody}$_3$, we choose the following representation for the solvent flux}
\begin{align}
\label{eq:flux3d}
 \mb{h} = -\mb{D}\nabla\mu\,, \quad \mb{D} = \frac{c\,D}{\mc{R}T}\mb{I}\,,
 \end{align}
 \rr{where $\mb{D}$ is the mobility tensor.} 

We then proceed with formulating the thermodynamics for the swelling material surface. In an isothermal setting, \rr{the free energy imbalance requires that}
\begin{align}
\label{eq:dissinsurfint}
\frac{\mbox{d}}{\mbox{d}t}\int_{\mc{P}_s}{\psi_s}   \leq \mc{W} + \Upsilon\,,
\end{align}
\rr{where $\psi_s$ is the surface Helmholtz free energy per unit area, and where}
\begin{align}
\label{eq:powers}
\mc{W} = \int_{\mc{P}_s}{\mb{f}_s\cdot\dot{\mb{u}}_s} + \int_{\partial\mc{P}_s}{\mb{t}_s\cdot\dot{\mb{u}}_s} + \int_{\mc{P}_s}{c_n\dot{\delta}}\,, && \Upsilon = - \int_{\partial\mc{P}_s}{\mu_s \mb{h}_s\cdot\mb{m}_s} + \int_{\mc{P}_s}{\mu_s q_s}\,,
\end{align}
\rr{are the mechanical and chemical power, respectively}.
\rr{The swelling constraint \eqref{eq:areaconstraint} is enforced by adding the term $-\int_{\mc{P}_s}{p_s(\dot{\delta} J_s + \delta \dot{J_s}- \Omega_s\dot{c}_s)}$ to the left hand side of eq.~\eqref{eq:dissinsurfint}, where $p_s$ is a Lagrange multiplier that represents the surface solvent pressure.}
Using the \rr{surface divergence theorem}, the balance equations \eqref{eq:balanceforces}-\eqref{eq:balathick} and \eqref{eq:localmassbalsurf}, and the continuity condition \eqref{eq:contcond}, eq.~\eqref{eq:dissinsurfint} may be localized as
\begin{align}
\label{eq:dissinsurf}
\dot{\psi}_s \leq \mb{S}^c_s\cdot\dot{\wh{\mb{F}}} + \sigma_n^c \dot{\delta} + \mu^c_s \dot{c}_s - \mb{h}_s\cdot\grads{\mu_s}\,,
\end{align}
where $\mb{S}_s^c = \mb{S}_s+p_s\delta\mb{P}_t^T\mb{F}_s^\star$, $\sigma^c_n = \sigma_n + p_s J_s$ and $\mu^c_s = \mu_s-p_s\Omega_s$ are the constitutively determinate parts of the surface stresses and of the surface chemical potential.
Inequality \eqref{eq:dissinsurf} suggests that the constitutive functions that deliver $\psi_s, \mb{S}^c_s, \sigma_n^c, \mu^c_s$ and $\mb{h}_s$ have to be prescribed. In particular, we assume that all these functions depend on $(\wh{\mb{F}}, \delta, c_s)$, while $\mb{h}_s$ also depends on $\grads{\mu}_s$.
By requiring that eq.~\eqref{eq:dissinsurf} be satisfied for every admissible constitutive process, we obtain the following thermodynamic restrictions
\begin{align}
&\mb{S}_s = \frac{\partial \psi_s}{\partial \wh{\mb{F}}}-p_s\delta\mb{P}_t^T\mb{F}_s^\star\,, && \sigma_n = \frac{\partial \psi_s}{\partial \delta}-p_s J_s\,, \label{eq:const1}  \\
&\mu_s = \frac{\partial \psi_s}{\partial c_s}+p_s\Omega_s\,, && \mb{h}_s(\wh{\mb{F}},\delta,c_s,\grads{\mu}_s) \cdot \grads{\mu}_s \leq 0\,. \label{eq:const3} 
\end{align}
Substitution of eq.~$\eqref{eq:const1}_2$ into the local form of the balance equation \eqref{eq:balathick} allows to obtain an expression for the pressure
\begin{align}
\label{eq:pb}
p_s = \frac{1}{J_s}\left(\frac{\partial \psi_s}{\partial \delta}-c_n\right)\,,
\end{align}
which can be used to eliminate $p_s$ from the governing equations of the model. Also notice that the microstructure $\delta$ can be eliminated too from the formulation, through eq.~\eqref{eq:areaconstraint}, and it is thus called a \textit{latent microstructure} \citep{Capriz1989}. Hence, we are left with $\mb{u}_s$ and $c_s$ as the primary unknowns of the governing equations for the material surface.

In analogy with eq.~\eqref{eq:flux3d}, we satisfy the requirement $\eqref{eq:const3}_2$ by choosing the following constitutive law:
\begin{align}
\label{eq:fluxsurf}
\mb{h}_s = - \frac{c_s D_s}{\mc{R}T}\grads{\mu_s}\,,
\end{align}
where $D_s$ is the diffusivity of the solvent within the boundary membrane.

\paragraph{Remark} Without the continuity condition \eqref{eq:contcond}, the last term in eq.~\eqref{eq:powers}$_2$ should be replaced by the contributions
\begin{align}
\int_{\mc{P}_s}{\mu_s q_e} - \int_{\mc{P}_s}{\mu|_{\mc{S}} q}\,,
\end{align}
where we have distinguished between the chemical potential $\mu|_{\mc{S}}$ associated to the flux $q$, and the chemical potential $\mu_s$ associated to the external source $q_e$. Then, \rr{from eq.~\eqref{eq:dissinsurf}, 
together with the constitutive restrictions} \eqref{eq:const1}-$\eqref{eq:const3}_1$ and \eqref{eq:fluxsurf}, we obtain the reduced free energy imbalance
\begin{align}
(\mu|_{\mc{S}}-\mu_s)\mb{h}|_{\mc{S}}\cdot\mb{m} \geq 0\,,
\end{align}
which may be satisfied, for instance, \rr{by imposing the Robin-like constraint}
\begin{align}
(\mu|_{\mc{S}}-\mu_s) = k(\mb{h}|_{\mc{S}}\cdot\mb{m})\,, \quad k \geq 0\,.
\end{align}
A similar discussion can be found in \cite{McBride2011}.

\subsection{A surface free energy for polymer gels}

In this section we focus on the derivation of a surface energy density for a membrane made of a polymer gel, whose dry shear modulus is $G_s$ and whose solvent-polymer interaction parameter is $\chi_s$. For such a swelling membrane, we take eq.~\eqref{eq:flory3d} as a representation form for the free energy, with $G$ replaced by $G_s$, $\lambda_{\rm o}$ by $\lambda_{s\rm o}$, $J_{\rm o}$ by $J_{s\rm o}=\lambda_{s\rm o}^3$ and $\chi$ by $\chi_s$. To reduce this the free energy to a surface energy density, we employ again the representation \eqref{eq:flayer} for \rr{the} deformation gradient of the membrane and the approximation $c \approx c|_{\mc{S}}$ for the concentration field, as done in Section \ref{sec:swellingconstraints}. 
The kinematic hypotheses on $\mb{F}$ and $c$ are limited, as in \cite{LibaiSimmonds1998}, to the derivation of the surface free energy. Then, by integrating eq.~\eqref{eq:flory3d} over the thickness, we get
\begin{align}
\psi_s(\wh{\mb{F}},c_s) = \int_{0}^{h}{\psi(\mb{F},c)} = \frac{1}{2}\frac{G_s}{J_{s\rm o}}h(\lambda_{s\rm o}^2\mbox{tr}(\mb{C}_s)+\lambda_{s\rm o}^2\delta^2-3) + \frac{1}{J_{s\rm o}}\frac{\mc{R}T}{\Omega_s} g(c_s) \,,
\end{align}
with
\begin{align}
g(c_s) = \Omega_s J_{s\rm o} c_s\log\left(\frac{\Omega_s J_{s\rm o} c_s}{1+\Omega_s J_{s\rm o} c_s}\right)+\chi_s \frac{\Omega_s J_{s\rm o} c_s}{1+\Omega_s J_{s\rm o} c_s}\,,
\end{align}
so that eqs.~\eqref{eq:const1}-$\eqref{eq:pb}$ yield
\begin{align}
&\mb{S}_s = \frac{G_s}{\lambda_{s\rm o}} h\wh{\mb{F}}-p_s\delta\mb{P}_t^T\mb{F}_s^\star = \frac{G_s}{\lambda_{s\rm o}} h\wh{\mb{F}}- p_s\frac{\delta}{J_s}\wh{\mb{F}}\mb{C}_s^\star\,, \label{eq:constsurfstress} \\
&p_s = \frac{G_s}{\lambda_{s\rm o}} h \frac{\delta}{J_s}-\frac{c_n}{J_s}\,, \label{eq:constsurfpress} \\
&\mu_s = \mc{R}T\left[\log{\frac{\Omega_s J_{s\rm o} c_s}{1+\Omega_s J_{s\rm o}  c_s}}+\frac{1}{1+\Omega_s J_{s\rm o} c_s} + \frac{\chi_s}{(1+\Omega_s J_{s\rm o} c_s)^2} \right] + \Omega_s p_s\,. \label{eq:constmusurf}
\end{align}

\paragraph{Free swelling equilibrium}

The free swelling equilibrium is attained when the system is allowed to swell without any applied loads ($\mb{t}=\mb{t}_s = \mb{f} = \mb{f}_e =\mb{0}$, $c_n=0$) or mechanical constraints and to attain chemical equilibrium with an external solvent whose chemical potential $\mu_e$ is fixed and homogeneous. The chemo-mechanical equilibrium is characterized by the conditions of zero stress and homogeneous chemical potential:
\begin{align}
\label{eq:freeswell}
\mb{S} = \mb{0}\,, && \mb{S}_s = \mb{0}\,, && \sigma_n = 0\,, && \mu=\mu_s=\mu_e\,,
\end{align}
so that, from the constitutive equations \eqref{eq:constbody} and \eqref{eq:constsurfstress}-\eqref{eq:constsurfpress}, it results that the deformation gradient  is isotropic and homogeneous, for both the body and the surface:\footnote{Here we assume that $\mb{g}_\alpha$ and $\mb{a}^\alpha$ are orthonormal bases.}
\begin{align}
\mb{F} = \lambda\mb{I}\,, && \mb{F}_s = \lambda_s\,\mb{g}_\alpha \otimes \mb{a}^\alpha\,, && \delta = \lambda_s\,,
\end{align}
with the pressure fields given by
\begin{align}
\label{eq:pressfreeswell}
p = \frac{G}{\lambda_{\rm o} \lambda} \,,  && p_s = \frac{G_s}{\lambda_{s\rm o}}\,\frac{h}{\lambda_s}\,.
\end{align}
The concentration fields, from \eqref{eq:swellingconstr} and \eqref{eq:areaconstraint}, are readily computed as
\begin{align}
\label{eq:concfreeswell}
c = c_{\rm o} + \frac{\lambda^3-1}{\Omega}\,, && c_s = c_{s\rm o} + \frac{\lambda_s^3-1}{\Omega_s}\,,
\end{align}
while 
the constitutive equations \eqref{eq:flux3d} and \eqref{eq:fluxsurf} \rr{imply that there is no solvent flow}: $
\mb{h} = \mb{0}$, $\mb{h}_s = \mb{0}$. By substituting expressions \eqref{eq:concfreeswell} and \eqref{eq:pressfreeswell} in $\eqref{eq:constbody}_2$ and \eqref{eq:constmusurf} we obtain a set of non-linear algebraic equations:
\begin{align}
\begin{split}
&\log{\frac{ (\lambda\lambda_{\rm o})^3 -1}{(\lambda\lambda_{\rm o})^3}}+\frac{1}{(\lambda\lambda_{\rm o})^3} + \frac{\chi}{(\lambda\lambda_{\rm o})^6} + \frac{G \Omega}{\mc{R}T}\,\frac{1}{\lambda\lambda_{\rm o}} = \frac{\mu_e}{\mc{R}T}\,, \label{eq:chemeq2} \\
&\log{\frac{ (\lambda_s\lambda_{s\rm o})^3 -1}{(\lambda_s\lambda_{s\rm o})^3}}+\frac{1}{(\lambda_s\lambda_{s\rm o})^3} + \frac{\chi_s}{(\lambda_s\lambda_{s\rm o})^6} + \frac{G_s \Omega}{\mc{R}T}\,\frac{1}{\lambda_s\lambda_{s\rm o}} = \frac{\mu_e}{\mc{R}T}\,,
\end{split}
\end{align}
which may be solved for the free swelling stretches $\lambda$ and $\lambda_s$. \rr{For $\lambda=\lambda_s=1$, these equations define the relations} between the dimensionless chemical potential $\mu_e/\mc{R}T$ of the external solvent and the initial swelling stretches $\lambda_{\rm o}$ and $\lambda_{s\rm o}$, depending on the dimensionless parameters $G\Omega/\mc{R}T$, $G_s\Omega/\mc{R}T$, $\chi$ and $\chi_s$. Notice that, when the body and the surface are made of the same material ($G=G_s$, $\chi = \chi_s$, which implies $\lambda_{\rm o} = \lambda_{s\rm o}$) $\lambda = \lambda_s$, as expected.

\section{Boundary conditions and weak form of the governing equations}
\label{sec:weakform}

As regards the mechanical boundary conditions, we assume that the boundary loads $\mb{t}$ and $\mb{t}_s$ may be assigned on the portions $\partial_\mb{t}\mc{B}$ and $\partial_\mb{t}\mc{S}$ of $\partial \mc{B} \setminus \mc{S}$ and $\partial \mc{S}$, respectively, while the displacements $\mb{u}$ and $\mb{u}_s$ may be prescribed on $\partial \mc{B}_{\mb{u}}$ and $\partial \mc{S}_{\mb{u}}$.
For solvent transport, the boundary sources $-\mb{h}\cdot\mb{m} = q$ and $-\mb{h}_s\cdot\mb{m}_s$ may be prescribed on $\partial_q\mc{B} \subset \partial \mc{B} \setminus \mc{S}$ and $\partial_q\mc{S} \subset \partial \mc{S}$, respectively; the chemical potentials $\mu$ and $\mu_s$ may be prescribed on $\partial_\mu\mc{B} \subset \partial \mc{B} \setminus \mc{S}$ and $\partial_\mu\mc{S} \subset \partial \mc{S}$, respectively. The latter boundary conditions correspond to the assumption of instantaneous chemical equilibrium between the solvent within the gel and the external solvent, and they can be considered as implicit Dirichlet boundary conditions for $c$ and $c_s$. Indeed, we may solve the non-linear algebraic equations $\mu=\mu_e$ and $\mu_s = \mu_e$ written in weak form as
\begin{align}
\label{eq:chempotcontweak}
\int_{\partial_\mu\mc{B}}{(\mu(\bar{c},p)-\mu_e)\tilde{c}}=0\,, && \int_{\partial_\mu\mc{S}}{(\mu_s(\bar{c}_s,p_s)-\mu_e)\tilde{c}_s}=0\,,
\end{align}
for the auxiliary unknowns $\bar{c}$ and $\bar{c}_s$, where $\mu$ and $\mu_s$ are given by $\eqref{eq:constbody}_2$ and \eqref{eq:constmusurf}, and then prescribe the solutions as essential boundary conditions for $c$ and $c_s$: $c = \bar{c}$ on $\partial_\mu\mc{B}$ and $c_s = \bar{c}_s$ on $\partial_\mu\mc{S}$. An analogous approach is used to impose  pointwise chemical equilibrium of the surface $\mc{S}$ with the external solvent.

We now summarize the governing equations that have been presented so far recast in weak form, with $(\tilde{\mb{u}},\tilde{p},\tilde{\mu},\tilde{\mu}_s,\tilde{g})$ the test fields corresponding to the unknowns $(\mb{u},p,c,c_s,g)$ of the problem:
\begin{itemize}
\item Balance of forces for the body and the boundary material surface
\begin{align}
&-\int_{\mc{B}}{\mb{S}\cdot\nabla\tilde{\mb{u}}}-\int_{\mc{S}}{\mb{S}_s\cdot\grads\tilde{\mb{u}}}+\int_{\partial_\mb{t}\mc{B}}{\mb{t}\cdot\tilde{\mb{u}}}+\int_{\mc{S}}{\mb{f}_e\cdot\tilde{\mb{u}}}+\int_{\partial_\mb{t}\mc{S}}{\mb{t}_s\cdot\tilde{\mb{u}}}=0\,; \label{eq:balforcesweak}
\end{align}
\item Swelling constraint for the body
\begin{align}
&\int_{\mc{B}}{[J-1-\Omega\,(c-c_{\rm o})]\tilde{p}}=0\,; \label{eq:volumeconstrweak}
\end{align}
\item Balance of solvent mass for the body
\begin{align}
&-\int_{\mc{B}}{(\dot{c}\tilde{\mu}-\mb{h}\cdot\nabla\tilde{\mu})}+\int_{\partial_q\mc{B}}{q\tilde{\mu}}+\int_{\mc{S}}{g \tilde{\mu}}=0\,; \label{eq:balsolbodyweak}
\end{align}
\item Balance of solvent mass for the boundary material surface
\begin{align}
&-\int_{\mc{S}}{[(\dot{c}_s+g-q_e)\tilde{\mu}_s-\mb{h}_s\cdot\grads\tilde{\mu}_s]}-\int_{\partial_q\mc{S}}{(\mb{h}_s\cdot\mb{m}_s)\tilde{\mu}_s}=0\,; \label{eq:balsolsurfweak}
\end{align}
\item Continuity of the chemical potential between the body and the surface
\begin{align}
\label{eq:chempotcontweak}
\int_{\mc{S}}{(\mu-\mu_s)\tilde{g}}=0\,.
\end{align}
\end{itemize}
The latter equation determines the solvent flux $g = -\mb{h}|_{\mc{S}}\cdot\mb{m}$. In writing equation~\eqref{eq:balforcesweak}, we have expressed the traction $\mb{t}$ on $\mc{S}$ as $\mb{t} = \mb{f}_e - \mb{f}_s = \mb{f}_e + \divgs{\mb{S}_s}$ by eq.~\eqref{eq:balanceforces}, and we have used the continuity of the displacement up to $\mc{S}$.
As initial conditions we prescribe:
\begin{align}
&\mb{u} = \mb{0}\,, \quad \mbox{on}\ \mc{B}\,, && \mb{u}_s = \mb{0}\,, \quad \mbox{on}\ \mc{S}\,, \\
&p = \frac{G}{\lambda_{\rm o}}\,, \quad \mbox{on}\ \mc{B}\,, && g = 0\,, \quad \mbox{on}\ \mc{S}\,,\\
&c = c_{\rm o}\,, \quad \mbox{on}\ \mc{B}\,, && c_s = c_{s\rm o}\,, \quad \mbox{on}\ \mc{S}\,, 
\end{align}
where the initial pressure $p$ is computed from eq.~\eqref{eq:pressfreeswell} with $\lambda = 1$, and $c_{\rm o} = (\lambda_{\rm o}^3-1)/(\Omega \lambda_{\rm o}^3)$, $c_{s\rm o} = (\lambda_{s\rm o}^3-1)/(\Omega_s \lambda_{s\rm o}^3)$ are defined by the initial swelling stretches $\lambda_{\rm o}$, $\lambda_{s\rm o}$.

We choose the same test functions for $c$ and $c_s$ so that, by summing eq.~\eqref{eq:balsolbodyweak} and eq.~\eqref{eq:balsolsurfweak} with $\tilde{\mu}=\tilde{\mu}_s$, we recover the weak form of the balance of solvent mass for the system body + material surface:
\begin{align}
-\int_{\mc{B}}{(\dot{c}\tilde{\mu}-\mb{h}\cdot\nabla\tilde{\mu})}+\int_{\partial_q\mc{B}}{q\tilde{\mu}} -\int_{\mc{S}}{[(\dot{c}_s-q_e)\tilde{\mu}-\mb{h}_s\cdot\grads\tilde{\mu}]}-\int_{\partial_q\mc{S}}{(\mb{h}_s\cdot\mb{m}_s)\tilde{\mu}}=0\,.
\end{align}
For the following applications, the weak form equations \eqref{eq:balforcesweak}-\eqref{eq:chempotcontweak}, together with the constitutive equations \eqref{eq:constbody}-\eqref{eq:flux3d}, \eqref{eq:fluxsurf}, and \eqref{eq:constsurfstress}-\eqref{eq:constmusurf}, the swelling constraint \eqref{eq:areaconstraint}, and the boundary conditions appropriate for the problem at hand, are implemented in the software COMSOL Multiphysics v4.4 and solved using the finite element method.

\section{Applications}
\label{sec:applications}
We present several numerical examples to demonstrate the applicability of the model to the study of biomedical devices, soft crawler robots and surface transport phenomena. 

\subsection{Drug release from a hydrogel disk}

Certain polymers, such as NIPAAm and its copolymers, exhibit a dramatic de-swelling when temperature is increased beyond a threshold. Based on these polymers, temperature-responsive hydrogels have been produced and investigated as drug delivery systems, where temperature acts as an external stimulus that modulates the drug release rate. For instance, Bae and coworkers \citep{Bae1991} fabricated drug-loaded disks made of PNIPAAm-PTMEG interpenetrating polymer networks, which realize an on-off, pulsated drug release. In particular, upon increasing the temperature from $298\ \mbox{K}$ to $303 \ \mbox{K}$, there was a 
shrinkage of the outer membrane leading to a surface layer with very low permeability to the drug that blocked further drug release. 

Inspired by this experiment, we consider a hydrogel disk where, for simplicity, only the coating layer consists of a thermo-responsive hydrogel.
As in the experiment, the disk has a diameter of $10\ \mbox{mm}$ and a thickness of $1\ \mbox{mm}$. Since the coating layer (boundary membrane) is very thin (thickness $h=1\ \mu\mbox{m}$), we neglect heat exchange and assume that thermal equilibrium is reached instantaneously, whenever temperature is changed. For the initial condition, we assume that the hydrogel disk is fully swollen, in equilibrium with pure water outside ($\mu_e = 0\ \mbox{J}/\mbox{mol}$), and that the contribution of the drug to swelling is negligible with respect to that given by water. The disk is constraint-free and its boundary is traction-free; bulk forces are absent. 

In the framework of Flory-Rehner swelling theory, thermo-responsive hydrogels are modeled by prescribing a dependence on temperature of the mixing affinity between polymer and solvent, represented by the dimensionless parameter $\chi$. In particular, following \cite{Chester2011}, the temperature dependence of the solvent-polymer interaction parameter $\chi_s$ for the coating layer is taken as
\begin{align}
\label{eq:chib}
\chi_s(T) = \frac{1}{2}(\chi_L + \chi_H)-\frac{1}{2}(\chi_L-\chi_H)\tanh\left( \frac{T-T_t}{\Delta T} \right)\,,
\end{align}
where $T_t$ is the transition temperature, $\chi_L$ ($\chi_H$) is the value of $\chi_s$ below (above) the transition temperature, and $\Delta T$ is the temperature change in the transition from $\chi_L$ to $\chi_H$.

\begin{figure}[!h]
\centering
\includegraphics[scale=1.1]{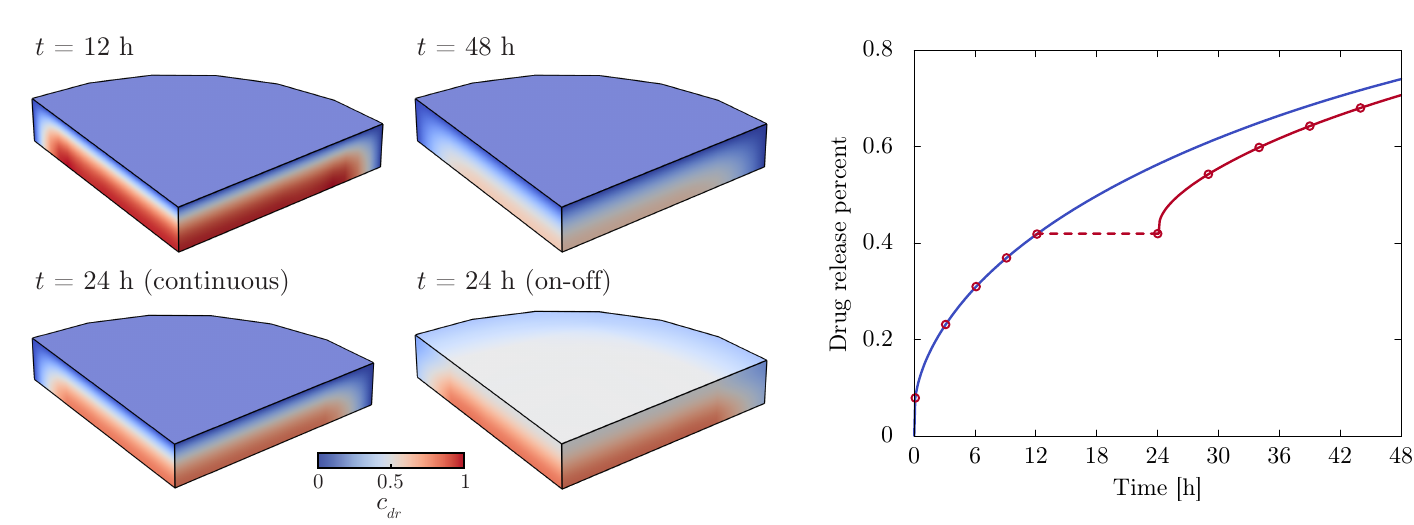}
\caption{Drug release from a hydrogel disk coated with a thermo-responsive gel. When temperature is increased beyond the transition threshold, the coating gel shrinks, causing a reduction in the permeability of the surface, which hampers further drug release. (Left) Contour plots of the normalized drug concentration $c_{dr}$ over 1/8 of the hydrogel disk at several times, for the continuous and on-off release modes. (Right) Percent of drug released from the disk as a function of time, for continuous (blue solid line) and on-off (off = dashed line, on = red solid line with circles) release.}
\label{fig:drug}
\end{figure}

As usual in diffusion-controlled drug delivery systems, where drug diffusion is the rate-limiting process, transport of drug within the hydrogel is described by the equation \citep{Siepmann2012}
\begin{align}
\dot{c}_{dr} = D_{dr}\Delta c_{dr}\,,
\end{align}
where $c_{dr}$ is the normalized drug concentration, which is initially homogeneous and equal to $1$, and $D_{dr}$ is the drug diffusivity coefficient within the hydrogel. When the membrane is swollen, its permeability increases because of the increased mesh size of the polymer network, thus allowing for a higher flux of solvent. To model the swelling-dependent permeability of the membrane, we describe the drug flux through it as
\begin{align}
\label{eq:drugflux}
h_{dr} = -D_{dr}\nabla c_{dr} \cdot \mb{m} = - k(J_s \delta)^n  (c_{dr}|_{\mc{S}}-c_{\rm ext})\,,
\end{align}
where $k$ is the permeability of the membrane to the drug when $J_s \delta = 1$, and $c_{\rm ext} = 0$ is the normalized drug concentration in the external medium (assuming perfect sink conditions). In the following, we set $n=10$. 

We performed transient numerical simulations of  two drug release profiles:
\begin{itemize}
\item continuous drug release: the temperature $T$ of the medium is held fixed at $T = 298\ \mbox{K} < T_t$;
\item on-off drug release: the temperature $T$ is held fixed at $298\ \mbox{K}$ for 12 hours, then it is instantaneously increased to $T = 303\ \mbox{K} > T_t$ and kept constant for 12 hours, and is finally decreased back to $T=298\ \mbox{K}$.
\end{itemize}
The values for the parameters used in numerical simulations are reported in Table~\ref{tabparam1}. During the simulations, we monitored the percent of drug released in the medium $\int_{\mc{B}}{(1-c_{dr})}/V$, where $V$ is the volume of the hydrogel disk.
In the continuous drug release case, the drug diffuses continuously out of the disk, while in the on-off drug release, the coating gel shrinks when temperature is increased beyond the transition threshold, causing a reduction in the permeability of the surface, which hampers further drug release (Figure~\ref{fig:drug}). During the off drug release period, the drug concentration in the hydrogel tends to equalize, so that  $c_{dr}$ increases at the boundary with respect to its value immediately after the step increase of temperature ($t = 12\ \mbox{h}$). As a result, when drug release is switched on again ($t = 24\ \mbox{h}$), the flux $h_{dr}$ given by eq.~\eqref{eq:drugflux} increases with respect to the value at drug release switch-off, as indicated by the corresponding slopes of the release profile in Figure~\ref{fig:drug}.

\begin{table}
\centering
\caption{Parameter values used in numerical simulations.}
\label{tabparam1}
\begin{tabular}{lcl}
\toprule
Parameter		&	Value  						      		&	Description \\
\midrule
$\Omega$		&	$6\times10^{-5}\,\mbox{m}^3/\mbox{mol}$ 	&	\mbox{Solvent molar volume} \\
$D$			&	$1\times10^{-9}\,\mbox{m}^2/\mbox{s}$ 		&	\mbox{Bulk solvent diffusivity} \\
$D_s$		&	$D$     								&	\mbox{Surface solvent diffusivity} \\
$D_d$		&	$2\times10^{-12}\,\mbox{m}^2/\mbox{s}$ 		&	\mbox{Bulk drug diffusivity} \\
$k$			&	$D_d/h$								&	\mbox{Drug permeability of the boundary membrane} \\
$G$			&	$10\ \mbox{kPa}$ 						&	\mbox{Shear modulus of the bulk gel} \\
$G_{s}$		&	$10\ \mbox{kPa}$						&	\mbox{Shear modulus of the boundary membrane}	\\
$\chi$		&	0.2									&	\mbox{Bulk solvent-polymer mixing parameter}				\\
$\chi_H$		&	0.6 									&	\mbox{Surface solvent-polymer mixing parameter below $T_t$}	\\
$\chi_L$		&	0.2 									&	\mbox{Surface solvent-polymer mixing parameter above $T_t$}	\\
$T_t$		&	$301\,\mbox{K}$						&	\mbox{Transition temperature for the thermo-responsive gel} \\
$\Delta T$		&	$0.5\,\mbox{K}$						& 	\mbox{Temperature interval of transition from $\chi_L$ to $\chi_H$} \\
$T$			&	$298\,\mbox{K}$ (on), $303\,\mbox{K}$ (off)  	&	\mbox{Ambient temperature} \\
$\lambda_{\rm o}$	&	1.5									&	\mbox{Initial swelling stretch of the bulk gel}  \\
$\lambda_{s\rm o}$ &	1.5 									&	\mbox{Initial swelling stretch of the boundary membrane} \\
\bottomrule
\end{tabular}
\end{table}

\subsection{Temperature-activated crawler}

As a prototype of a temperature-activated crawler, we consider a hydrogel beam whose bottom surface is coated with a thin layer consisting of a thermo-responsive gel. The beam is $L=20\ \mbox{mm}$ long, $b=1\ \mbox{mm}$ wide and $h_{\rm o} = 2\ \mbox{mm}$ thick, including the thickness of the coating layer, which is $\beta = 1/10$ of the total thickness. The material parameters are taken as in Table~\ref{tabparam1}. The crawler interacts with a directional substrate \citep{Hancock2012} that exerts a friction force only sensitive to the sign of the sliding velocity, that is, a directional dry-friction interaction \citep{Gidoni2014, Noselli2014, DeSimone2015}. In particular, the friction force for positive velocity is less than for negative velocity. The crawler is initially in free-swelling equilibrium with a pure liquid solvent ($\mu_e = 0\ \mbox{J}/\mbox{mol}$); the free-swelling stretches of the beam and the coating layer with respect to the dry configurations are $\lambda_{\rm o}$ and $\lambda_{s\rm o}$, respectively. For a time interval $\tau$, we prescribe a temperature profile that linearly increases from $T_{\rm o} = 298\ \mbox{K}$ to $T_m = 300\ \mbox{K}$, which corresponds to an increase in $\chi_s$  from $0.2$ to $0.4$ in the half-cycle $\tau/2$, and then linearly decreases to $T_{\rm o}$ in the remaining half-cycle. As for the previous application, we assume that thermal equilibrium is instantaneous, thus neglecting transient heat transfer.

\begin{figure}[!h]
\centering
\includegraphics[scale=1.1]{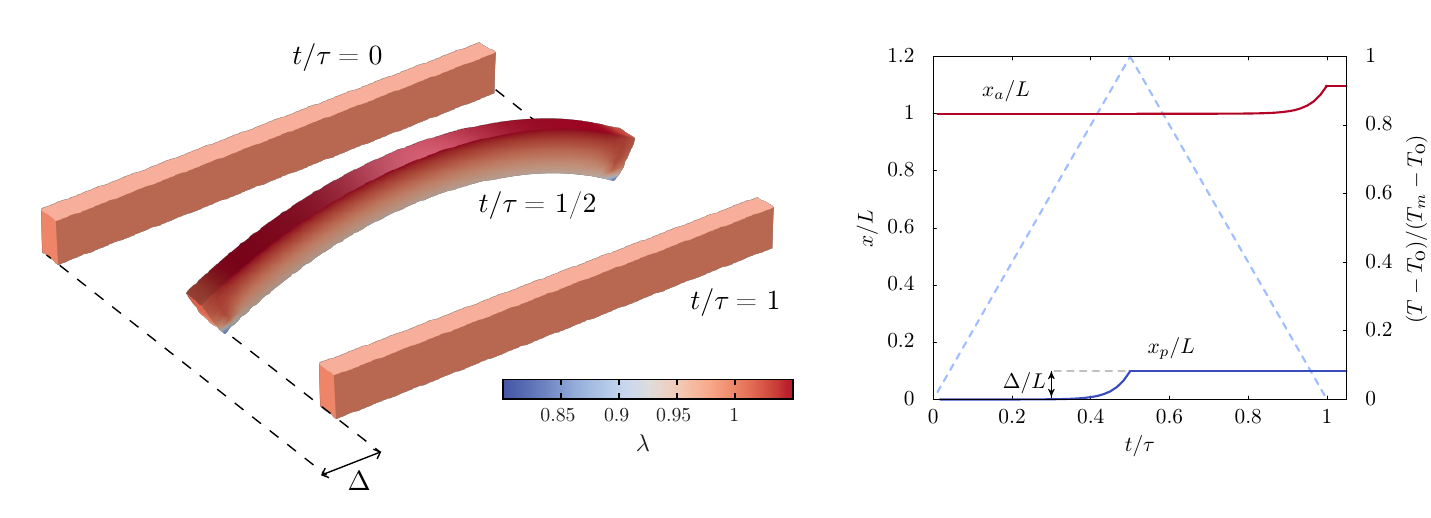}
\caption{Motion of the temperature-activated crawler. (Left) Snapshots of the crawler advancing on a directional substrate by exploiting dry friction. When temperature is increased from $T_{\rm o}$ to $T_m$, the crawler bends and the posterior edge in contact with the surface moves over it, while the anterior edge stays fixed. Color code represents the longitudinal stretch $\lambda$ of the crawler. (Right) Dimensionless positions $x_p/L$ and $x_a/L$ of the posterior and anterior edges, respectively, as a function of the scaled time $t/\tau$ during a period $\tau$ corresponding to one temperature cycle. The dashed line is the temperature profile.}
\label{fig:crawler}
\end{figure}

When temperature is increased from $T_{\rm o}$ to $T_m$, the coating layer shrinks, as solvent is expelled due to the reduction in the solvent-polymer chemical affinity, and the crawler bends. Because of the directional interaction with the substrate, the posterior edge of the beam that is in contact with the surface slides over it, while the anterior edge sticks to the substrate (Figure~\ref{fig:crawler}). As temperature is decreased to $T_{\rm o}$, the crawler returns to its straight configuration as the coating layer swells, with the anterior edge sliding on the substrate and the posterior edge staying fixed. The sliding $\Delta$ of the posterior edge in the first half of the cycle equals the advancement of the crawler in one temperature cycle, which is about $L/10$ for the crawler considered here.

\begin{figure}[!h]
\centering
\includegraphics[scale=1.1]{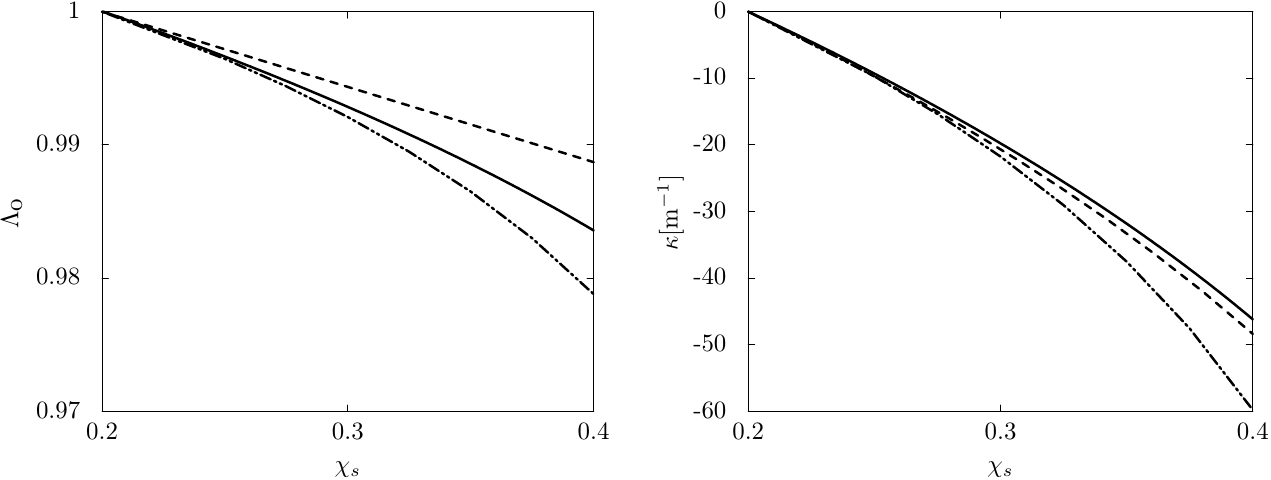}
\caption{Comparison among the results obtained with several models of the crawler. Longitudinal stretch $\Lambda_{\rm o}$ (left) and curvature $\kappa$ (right) of the longitudinal axis of the crawler as a function of the dimensionless mixing parameter $\chi_s$. The dashed-dotted line is the result obtained with the analytical bilayer beam model, the solid line is the result obtained with the fully three-dimensional model, while the dashed line is the result obtained with the swelling material surface model.}
\label{fig:compmemb3d}
\end{figure}

The longitudinal stretch and curvature of the longitudinal axis of the crawler can be estimated using a plane-bending model of a bilayer beam, as in \cite{Lucantonio08112014}. Under the assumption of plane cross-sections, we represent the longitudinal stretch of the beam as:
\begin{align}
\lambda(z) = \Lambda_{\textrm{o}}(1+z\Lambda_{\textrm{o}}\kappa)\,,
\end{align}
where $z \in [-h_{\rm o}/2,h_{\rm o}/2]$ is the thickness coordinate, $\Lambda_{\rm o}$ and $\kappa$ are the (uniform) stretch and curvature of the longitudinal axis $z=0$, respectively. We assume that the longitudinal stress in the beam is related to the difference between the actual stretch $\lambda$ and the stretches $\lambda^f$ and $\lambda^f_{s}$ that each layer would attain if it was free to swell separately from the other layer. Hence, we employ the following simplified constitutive equations for the longitudinal stresses within the layers
\begin{align}
&\sigma(z) = 3G\left(\frac{\lambda(z)}{\lambda^f}-1\right)\,, && \left(-\frac{h_{\rm o}}{2}+\beta h_{\rm o}\right)<z<\frac{h_{\rm o}}{2}\,, \\
&\sigma_s(z) = 3G_s\left(\frac{\lambda(z)}{\lambda^f_{s}}-1\right)\,, && -\frac{h_{\rm o}}{2}<z<\left(-\frac{h_{\rm o}}{2}+\beta h_{\rm o}\right)\,,
\end{align}
where $\lambda^f$ and $\lambda^f_{s}$ are determined by the free-swelling equilibrium eq.~\eqref{eq:chemeq2}:
\begin{align}
&\log{\frac{ (\lambda^f\lambda_{\rm o})^3 -1}{(\lambda^f \lambda_{\rm o})^3}}+\frac{1}{(\lambda^f \lambda_{\rm o})^3} + \frac{\chi}{(\lambda^f \lambda_{\rm o})^6} + \frac{G \Omega}{\mc{R}T}\,\frac{1}{\lambda^f \lambda_{\rm o}} = \frac{\mu_e}{\mc{R}T}\,, \\
&\log{\frac{ (\lambda_s^f \lambda_{s\rm o})^3 -1}{(\lambda_s^f \lambda_{s\rm o})^3}}+\frac{1}{(\lambda_s^f \lambda_{s\rm o})^3} + \frac{\chi_s}{(\lambda_s^f \lambda_{s\rm o})^6} + \frac{G_s \Omega}{\mc{R}T}\,\frac{1}{\lambda_s^f \lambda_{s\rm o}} = \frac{\mu_e}{\mc{R}T}\,.
\end{align}
By imposing that the  force and moment resultants over the cross-section of the crawler vanish, 
\begin{align}
&b\int_{-h_{\rm o}/2}^{-h_{\rm o}/2+\beta h_{\rm o}}{\sigma_s(z)} + b\int_{-h_{\rm o}/2+\beta h_{\rm o}}^{h_{\rm o}/2}{\sigma(z)} = 0\,, \\
&b\int_{-h_{\rm o}/2}^{-h_{\rm o}/2+\beta h_{\rm o}}{z\sigma_s(z)} + b\int_{-h_{\rm o}/2+\beta h_{\rm o}}^{h_{\rm o}/2}{z\sigma(z)} = 0\,,
\end{align}
we obtain the following system of equations, with $\Lambda_1 = \kappa \Lambda_{\rm o}^2$,
\begin{align}
&\left(A\frac{G}{\lambda^f}+A_s\frac{G_s}{\lambda^f_{s}}\right)\Lambda_{\rm o} + \left(S\frac{G}{\lambda^f}+S_s\frac{G_s}{\lambda^f_{s}}\right)\Lambda_1 = AG+A_s G_s\,, \\
&\left(S\frac{G}{\lambda^f}+S_s\frac{G_s}{\lambda^f_{s}}\right)\Lambda_{\rm o} +  \left(I\frac{G}{\lambda^f}+I_s\frac{G_s}{\lambda^f_{s}}\right)\Lambda_1 = SG+S_s G_s\,,
\end{align}
where $A, A_s$ are the areas of the cross-sections of the layers, $S, S_s$ are the static moments of such cross-sections, and $I, I_s$ are the moments of inertia.
These equations, together with the free-swelling equations, provide the longitudinal stretch $\Lambda_{\rm o}$ and the curvature $\kappa$ of the axis of the crawler. In Figure~\ref{fig:compmemb3d} we report the comparison among the solution of this system for different values of $\chi_s$, the numerical solution of the model where both the layers are treated as three-dimensional gels, and the numerical solution of the crawler model where the coating layer is modeled as a swelling material surface. We notice that, for moderate curvatures corresponding to values of $\chi_s \leq 0.4$, the agreement among the models is good, especially as regards the values of $\Lambda_{\rm o}$, which stay within a 1\% variation.

\subsection{Spreading and absorption of a liquid on the surface of a gel layer}

Surfaces of bodies usually display mechanical and diffusion properties that are different from those of bulk materials, thus affecting surface transport processes. In particular, spreading and absorption phenomena over porous surfaces are important for many technological applications, such as high speed inkjet printing on coated papers \citep{Kettle2010}. Here we study an example problem where the surface of a gel layer (edge length $L=100\ \mu\mbox{m}$, thickness $L/8$, volume $V$) has a different permeability to the solvent with respect to that of the bulk. The surface is modeled as a swelling material surface, with a thickness $h=1\ \mu\mbox{m}$. A constant solvent flux is prescribed for $\tau = 1\ \mbox{s}$ on a circular region with radius $10\ \mu\mbox{m}$ at the center of the surface. Over the period $\tau$, the total amount of solvent absorbed by the layer equals $V/8$. For the material parameters, we take $G = G_s = 40\ \mbox{kPa}$, $\Omega = 6 \times 10^{-5}\ \mbox{m}^3/\mbox{mol}$, $\chi = \chi_s = 0.2$, $T = 298\ \mbox{K}$, $D=10^{-10}\ \mbox{m}^2/\mbox{s}$. The boundary diffusivity $D_s$ varies between $D$ and $10^3 D$. 

\begin{figure}[!h]
\centering
\includegraphics[scale=1.1]{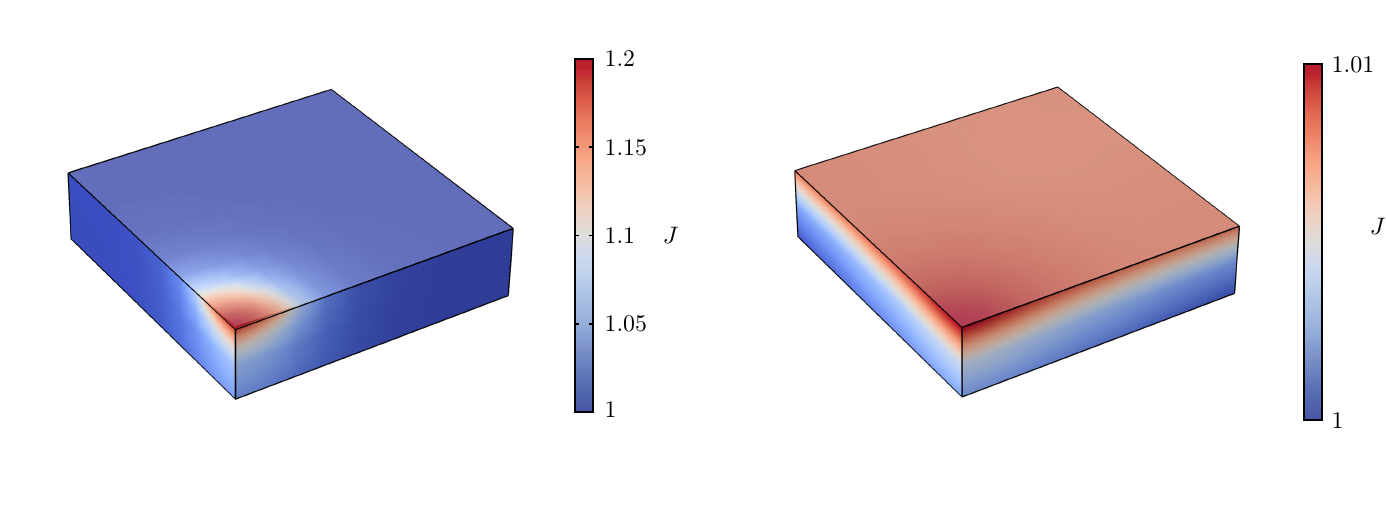}
\caption{A gel layer coated with a surface having a different solvent permeability. A solvent flux is prescribed for $1\ \mbox{s}$ in a circular region at the center of the layer. The solvent partially spreads over the surface, while it is absorbed by the layer. Plot of the swelling ratio $J$ at time $t = 1\ \mbox{s}$ over 1/4 of the gel layer for (left) $D_s = 1\times10^{-10}\ \mbox{m}^2/\mbox{s} = D$ and for (right) $D_s = 1\times10^{-7}\ \mbox{m}^2/\mbox{s} = 10^3 D$.}
\label{fig:spread}
\end{figure}

The results of the numerical simulations corresponding to different values of $D_s$ are reported in Figure~\ref{fig:spread}. When $D=D_s$, solvent absorption, \textit{i.e.}~migration along the thickness, is faster than surface spreading, \textit{i.e.}~in-plane transport, because the diffusion time scale is smaller along the thickness. As $D_s$ increases, surface spreading prevails on absorption, so that the solvent profile (and, thus, the swelling ratio $J$) tends to homogenize in plane faster than along the thickness.

\section{Conclusions}

In this paper, we have established a thermodynamically consistent theory for swelling material surfaces that allows to describe coupled elasticity and solvent transport in polymer gel membranes. The balance equations for a swelling material surface have been obtained from a virtual work functional, following a direct approach. A kinematical swelling constraint and a surface energy for the swelling material surface consistent with the Flory-Rehner theory have been derived. The governing equations of the model have been recast in weak form, thus allowing for their implementation in a finite element code.

We have applied the theory to the study of several model problems motivated by technological applications in the fields of biomedicine, micro-motility and coating technology. Specifically, we have performed numerical simulations of a smart drug delivery system, a temperature-activated micro-crawler and a coated gel layer subject to surface solvent spreading and absorption. We believe that the present theory may be employed as an effective design tool in such contexts.

\section*{Acknowledgments}
AL acknowledges support of Regione Friuli Venezia Giulia through Fondo Sociale Europeo~--~S.H.A.R.M. project and thanks Giovanni Noselli for useful discussions. LT acknowledges support of GNFM-INdAM (National Group of Mathematical Physics), Italy. ADS acknowledges support of the European Research Council through the Advanced Grant 340685~--~MicroMotility.


\end{document}